\documentclass[prl,twocolumn,epsf,psfig]{revtex4}

\def\widetext{\end{twocolumn}}
\input psfig
\input epsf
\input{tcilatex}

\begin{document}

\title{Theory for spin and orbital orderings in high temperature
phases in $YVO_3$}
\author{Theja. N. De Silva$^{a,b}$ , Anuvrat Joshi$^{c}$, Michael
Ma$^{a,d}$, and Fu
Chun Zhang$^{a,e}$}

\affiliation{$^{a}$Department of Physics, University of Cincinnati, 
Cincinnati Ohio 45221 \\
$^{b}$Department of physics, University of Ruhuna, Matara, Sri Lanka  \\
$^{c}$ National High Magnetic Field Laboratory,
Florida State University, Tallahassee, FL 32310\\
$^{d}$ Department of Physics, Chinese University of Hong Kong, 
Shatin, Hong Kong\\
$^{e}$ Institute of Theoretical Physics, Chinese Academy of Sciences, 
Beijing, China}

\begin{abstract}
\noindent Motivated by the recent neutron diffraction experiment on $YVO_3$,
we consider a microscopic model where each $V^{3+}$ ion is occupied by two $%
3d$ electrons of parallel spins with two fold degenerate orbital
configurations. The mean field classical solutions of the spin-orbital
superexchange model predicts an antiferro-orbital ordering at a higher
temperature followed by a C-type antiferromagnetic spin ordering at a lower
temperature. Our results are qualitatively consistent with the observed
orbital phase transition at $\sim 200$K and the spin phase transition at $%
\sim 114$K in $YVO_3$.
\end{abstract}

\maketitle

\section{I. \, Introduction}

The transition metal perovskite oxides exhibit many interesting physical
phenomena. In some of these compounds, the orbital degrees of freedom play
an important role in their magnetic properties due to the strong
spin-orbital coupling~\cite{mott,tokura,imada}. Examples include the
Mott-Hubbard type insulators $YVO_{3}$ and $LaVO_{3}$, which show very
unusual magnetic properties. Although the early experiments on $YVO_{3}$ and
$LaVO_{3}$ were reported back in the mid 1970's~\cite{boruk,zubkov1,zubkov2}%
, there has been renewed interest in the past decade on these materials ~%
\cite{kawano,nguyen,mahajan,kikuchi,cin,corti,ren,miyasaka,blake}. There are
two magnetic phases in $YVO_{3}$: C-type antiferromagnetic order
(ferromagnetic chains along the z-axis which stagger within the x-y plane)
at temperature $114$K $>T>77$K, and G-type antiferromagnetic order
(staggered in all three directions) at temperature $T<77$K~\cite
{boruk,zubkov1,zubkov2,kawano}. The magnetic order in $LaVO_{3}$ is always
C-type. The microscopic mechanism leading to the difference between these
two compounds is still under investigation, and it might be related to the
fact that at room temperature the cubic crystal structure is significantly
distorted in $YVO_{3}$
but almost undistorted in $LaVO_{3}$. It is generally believed that the
relevant orbital degrees of freedom, the degenerate or almost degenerate $%
3d-t_{2g}$ states are crucial to the observed magnetic properties.

There have also been interesting theoretical studies related to these
magnetic behaviors~\cite
{mizokawa1,mizokawa2,sawada,giniyat,shen,castellani,khomskii,kugel,euro}. In
particular, Khaliullin $\mathit{et.al}$~\cite{giniyat} considered a
spin-orbital Hamiltonian starting with 3-fold degenerate $t_{2g}$ orbitals,
and compared the free energies between the C-type and G-type spin states in $
YVO_3$ by including an explicit Jahn-Teller energy in the model.

Very recently, Blake $\mathit{et.al}$~\cite{blake} reported neutron
diffraction experiment in $YVO_3$, which shows clear evidence that the
orbital ordering has a sudden change from high temperature G-type to low
temperature C-type at the $77$K magnetic phase transition, manifested by a
change in the Jahn-Teller type of distortion. The data also show clear
evidence for the orbital transition from high temperature disordered phase
to the G-type ordered phase at $\sim 200$K. This has motivated us to study
the spin-orbital ordering in $YVO_3$.

In this paper, we consider a microscopic model for insulating $YVO_{3}$,
where each V-ion has two electrons with parallel spins favored by the large
Coulomb repulsion and the Hund's coupling. The distortion in the cubic crystal
structure already present at room temperature splits the degeneracy of the
  three $t_{2g}$
orbitals so that the $d_{xy}$ orbital is favored by the crystal 
field.  In our model,
we take it as always
singly occupied , while the other electron occupies either the 
$d_{xz}$ or the $
d_{yz}$ state. This description is consistent with the neutron diffraction
experiment~\cite{blake}. We consider the superexchange interaction of the
model and derive an effective Hamiltonian for $YVO_{3}$. We then study the
mean field classical solutions of the model, and examine the spin and
orbital orderings. We find a G-type orbital ordering at a higher temperature
followed by an additional C-type spin ordering at a lower temperature. Our
result is consistent with the observed orbital phase transition at $\sim 200$
K, and spin phase transition at $114$K in $YVO_{3}$. In this scenario, the
orbital ordering at $\sim 200$K is of the electronic origin, and the lattice
distortion at $\sim 200$K observed in the experiment is a consequence of the
orbital ordering and the electron-lattice coupling. The superexchange
interaction alone considered in our model does not explain the phase
transition at $77$K, which may require other interactions such as the
Jahn-Teller effect as proposed in previous articles~\cite{mizokawa2,giniyat}.

This paper is organized as follows. In Section II, we examine a multi-band
Hubbard model at electron density 2 electron per site, and consider the
limit of large Coulomb repulsion and the large Hund's coupling. We then
derive an effective Hamiltonian based on the superexchange mechanism. In
Section III, we discuss the mean field classical solutions of the model, and
examine the phase diagram for the orbital and spin orderings. A brief
summary is given in Section IV.

\section{ II.\, Model}

In $YVO_{3}$, the vanadium electron configuration is $3d^{2}$. The compound
has a cubic crystal structure, and each $V$ ion is surrounded by six oxygen
ions. Due to the cubic crystal field, the five-fold degenerate $3d$ orbitals
are split into a higher energy doublet of $e_{g}$ orbitals and a lower
energy triplet $t_{2g}$ orbitals. At low temperatures and for low energy
physics, the relevant orbitals are the three-fold $t_{2g}$ orbitals: $d_{xy}$%
, $d_{yz}$,$d_{zx}$. In the strong coupling limit, the on-site Coulomb
repulsion between the two electrons in the $3d$ states and the Hund's
coupling are much larger than the intersite electron hopping amplitudes, the
system is a Mott insulator with each V-ion having two localized electrons of
parallel spins in two out of three degenerate $t_{2g}$ orbitals. This
scenario appears to be consistent with experiments.

As indicated in the recent diffraction experiment~\cite{blake}, the cubic
crystal is distorted at room temperature. As a result, the V-O bond
distances are anisotropic. Here we consider the structure at room
temperature, where the V-O bond distance along $c$-axis (perpendicular axis)
is the smallest(see figure 1). This crystal structure further splits the $%
t_{2g}$ states. The $d_{xy}$ orbital has a lower energy, and becomes always
singly occupied. The other d-electron is either in $d_{yz}$ or $d_{zx}$
orbital. In the diffraction experiment~\cite{blake}, the data also indicate
a smaller difference in V-O bond lengths in the $xy$ plane, which we shall
neglect here for simplicity.

\begin{figure}[htb]
\epsfxsize=8.5cm \centerline{\epsffile{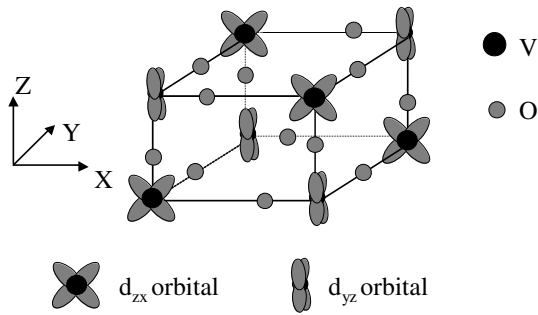}}
\caption{Idealized crystal structure for $YVO_{3}$ studied in this paper. $
V-O$ bond distance along the $z$-direction is shorter. The shown orbital
represent configurations $d_{xy}d_{yz}$ or $d_{xy}d_{xz}$.}
\label{}
\end{figure}

The atomic Hamiltonian~\cite{castellani} is then given by $H_0 = \sum_i H_i$
, where the sum over $i$ runs all the V-sites, and

\begin{eqnarray}
H_{i} &=&\frac{1}{2}\sum_{mm^{\prime },\sigma \sigma ^{\prime }}(1-\delta
_{mm^{\prime }}\delta _{\sigma \sigma ^{\prime }})U_{mm^{\prime
}}n_{im\sigma }n_{im^{\prime }\sigma ^{\prime }}  \nonumber \\
&&-J\sum_{mm^{\prime },\sigma }\biggr( n_{im\sigma }n_{im^{\prime }\sigma
}+c_{im\sigma }^{\dagger }c_{im-\sigma }c_{im^{\prime }-\sigma }^{\dagger
}c_{im^{\prime }\sigma }  \nonumber \\
&&-c_{im^{\prime }-\sigma }^{\dagger }c_{im^{\prime }\sigma }^{\dagger
}c_{im\sigma }c_{im-\sigma }\biggr) +\sum_{m,\sigma }\Delta _{m}n_{m\sigma }.
\end{eqnarray}

In the above Hamiltonian, $c_{im\sigma }^{\dagger }(c_{im\sigma })$ creates
(annihilates) an electron of orbital $m$ and spin $\sigma $ at site $i$, $%
n_{im\sigma }=c_{im\sigma }^{\dagger }c_{im\sigma }$. $\Delta _{1}=\Delta
_{2}=0$, and $\Delta _{3}=\Delta <0$, with $m=1,2,3$ representing orbitals $%
d_{yz}$, $d_{zx}$, $d_{xy}$ respectively. $U_{mm^{\prime }}$ is the on-site
direct interaction, and J is the exchange interaction, or the Hund's
coupling. For the $t_{2g}$ orbitals, $U_{mm}=U=U_{mm^{\prime }}+2J$ for $%
m^{\prime }\neq m$. In the case $U,J>\Delta $, this Hamiltonian leads to an
atomic ground state with each $V-3d^{2}$ ion having a total spin $S=1$ with
two-fold degenerate orbital configurations ($d_{xy},d_{xz}$) and ($%
d_{xy},d_{yz}$). This last restriction in orbital configurations is valid for
$YVO_3$ with strong lattice distortion but not for $LaVO_3$ where the 
cubic structure is almost undistorted at
room temperature.

We next introduce the intersite hopping Hamiltonian $H_{t}$, given by

\begin{eqnarray}
H_{t}=\sum_{\langle ij\rangle }\sum_{mm^{\prime },\sigma }\biggr(%
t_{m,m^{\prime }}^{ij}c_{i,m,\sigma }^{\dagger }c_{j,m^{\prime },\sigma
}+h.c.\biggr)
\end{eqnarray}

\noindent where the sum runs over all the nearest neighbor V-V pairs, and $%
t_{m,m^{\prime }}^{ij}$ is the electron hopping integral between two sites $%
i $ and $j$ from orbital $m$ to orbital $m^{\prime }$. Since the most
important contribution to the hopping integrals is from the path via the $2p$
state of the O-ion between the two neighboring V-ions, the hopping integrals
are diagonal in the present problem due to the cubic symmetry. Namely, we
have $t_{m,m^{\prime }}^{ij}=t_{m}^{ij}\delta _{m,m^{\prime }}$. Therefore,
there are only two independent hopping parameters, $%
t_{11}^{z}=t_{22}^{z}=t_{\perp }$, and $%
t_{22}^{x}=t_{33}^{x}=t_{11}^{y}=t_{33}^{y}=t_{\parallel }$, with the
super-index indicating the direction of the two sites. In the limit $
t_{\perp },t_{\parallel }<<U,\,J,\,\Delta $, the system is an insulator with
spin 1 on each V-ion. However, the virtual hopping introduces an effective
intersite coupling of spins and the occupied orbitals. The effective
Hamiltonian for $H=H_{0}+H_{t}$ can be derived by applying perturbation
theory to second order in $t_{\perp }$ or $t_{\parallel }$.

Let $|\phi _{ij}\rangle =|s_{i}^{z},\tau _{i}^{z},s_{j}^{z},\tau
_{j}^{z}\rangle $ be a ground state of $H_{0}$ for two V-ions $i,j$, where $%
s^{z}=1,\,0,\,-1$ is the spin z-component, and $\mathbf{\tau }$ is a
pseudospin-1/2 operator for the orbitals: $\tau ^{z}=1/2$ if $d_{yz}$ is
occupied, and $\tau ^{z}=-1/2$ if $d_{xz}$ is occupied. The matrix 
elements between the  unperturbed
ground states of the two V-ions can be calculated within the second order
perturbation theory, and it is given by,

\begin{eqnarray}
\langle \phi_{kl}| H_{eff} |\phi_{ij} \rangle = \sum_{I} \frac{\langle
\phi_{kl}|H_t| I \rangle\langle I | H_t |\phi_{ij}\rangle}{E_0 - E_I}
\end{eqnarray}

\noindent where the sum is over all the intermediate eigenstates $| I 
\rangle$ of $H_0$
corresponding to the eigen energy $E_I$, and $E_0$ is the ground state
energy of $H_0$. Two-electron states
with total spin $S=1$ are given in ref. 21. The electronic configuration 
of the intermediate state $| I
\rangle$ is $3d^3$ on one V-ion and $3d^1$ on the other. In the Appendix, we
list all the states for $V-3d^3$, and the corresponding energy difference $%
E_I - E_0$. The effective Hamiltonian can be derived from these matrix
elements. Defining for each site a  spin-1 operator $\mathbf{S}$ and a
pseudospin-1/2 operator $\mathbf{\tau }$ that act on the $s^z$ and 
$\tau^z$ degrees of freedom , it
can then be expressed as below,

\begin{eqnarray}
H_{eff}=\sum_{\left\langle ij\right\rangle ,\nu }[I^{\nu }(\mathbf{\tau }
_{i},\mathbf{\tau }_{j})\mathbf{S}_{i}\cdot \mathbf{S}_{j}+L^{\nu }(\mathbf{
\tau }_{i},\mathbf{\tau }_{j})]
\end{eqnarray}

\noindent where $\nu =x,y,z$ gives the direction of the bond $\left\langle
ij\right\rangle $. The first term corresponds to spin-dependent orbital
couplings while the second corresponds to orbital couplings which are spin
independent. The first term also shows that the effective spin-spin
couplings depend on orbital configuration. Equivalently, by defining $I^{\nu
}=K_{+}^{\nu }+K_{-}^{\nu }$ and $L^{\nu }=K_{+}^{\nu }-K_{-}^{\nu 
}$, $H_{eff}$ can be written as

\begin{eqnarray}
H_{eff}=\sum_{\left\langle ij\right\rangle ,\nu }[K_{+}^{\nu }(\mathbf{\tau }%
_{i},\mathbf{\tau }_{j})(\mathbf{S}_{i}\cdot \mathbf{S}_{j}+1) \\ \nonumber
+K_{-}^{\nu }(%
\mathbf{\tau }_{i},\mathbf{\tau }_{j})(\mathbf{S}_{i}\cdot \mathbf{S}%
_{j}-1)]
\end{eqnarray}

\noindent so that $2K_{+}^{\nu }$ and $2K_{-}^{\nu }$ are interpreted 
as the intersite
orbital couplings for parallel spins ($s_{i}^{z}=1$, $s_{j}^{z}=1$) and
antiparallel spins ($s_{i}^{z}=1$, $s_{j}^{z}=-1$) respectively. We choose
the energy unit to be $t_{\parallel }^{2}/U$, and denote $\eta =J/U$, $\eta
_{\frac{3}{2}}=1/(1-3\eta )$, $\eta _{\frac{1}{2}}=1/(1+2\eta )$, and $%
Q=t_{\perp }/t_{\parallel }$. $K_{\pm }^{\nu }$ can be expressed in terms of
parameters $Q$ and $\eta $, and are given below.

\begin{eqnarray}
&&K_{+}^{(x,y)}=\eta _{\frac{3}{2}}(\tau _{iz}\tau _{jz}-\frac{1}{4}),
\nonumber \\
&&K_{+}^{z}=2Q^{2}\eta _{\frac{3}{2}}(\vec{\tau}_{i}\cdot \vec{\tau}_{j}-%
\frac{1}{4}),  \nonumber \\
&&K_{-}^{(x,y)}=\alpha (\tau _{iz}\tau _{jz}-\frac{1}{4})+\frac{3}{4}(1+\eta
_{\frac{1}{2}})  \nonumber \\
&&+\frac{l_{x,y}}{4}(1+\eta _{\frac{1}{2}})(\tau _{iz}+\tau _{jz}),
\nonumber \\
&&K_{-}^{z}=Q^{2}\{2\alpha (\tau _{iz}\tau _{jz}-\frac{1}{4})+\frac{1}{2}%
(1+\eta _{\frac{1}{2}})  \nonumber \\
&&-\frac{1}{3}(\eta _{\frac{3}{2}}-1)(\tau _{i}^{+}\tau _{j}^{-}+\tau
_{i}^{-}\tau _{j}^{+})  \nonumber \\
&&-\frac{1}{2}(1-\eta _{\frac{1}{2}})(\tau _{i}^{+}\tau _{j}^{+}+\tau
_{i}^{-}\tau _{j}^{-})\}.
\end{eqnarray}

In the above equations, $\alpha =-\frac{1}{6}(1+2\eta _{\frac{3}{2}}-3\eta _{%
\frac{1}{2}})$, and $l_{x,y}=-1$ and $+1$ respectively. For a bond in the $z$
direction, where the $d_{xy}$ orbital is inert due to zero hopping amplitude
and the $d_{xz}$ and $d_{yz}$ hopping is isotropic, our model is similar to
the original Kugel-Khomskii model\cite{khomskii} with two differences. The
first is the replacement of spin $1/2$ by spin $1.$ The second is the effect
of the $d_{xy}$ occupation which changes the Hund's coupling contribution to
the on-site energies.

We first discuss the intersite pseudospin couplings between two parallel
spins. In this case, the psuedospin has a $SU(2)$ symmetry along $z$
-direction. Along $x$- or $y$-direction, however, the virtual hopping
integral for orbital $2$ or orbital $1$ vanishes, so there is no exchange
term in the pseudospin, and $K^{(x,y)}_+$ is of the Ising form. The
pseudospin coupling between the two V-ions of antiparallel spins is quite
different. There is a linear term $(\tau_{iz} + \tau_{jz})$ along $x$- or $y$%
-direction, which either favors $d_{zx}$ or $d_{yz}$ orbital occupation to
gain energy via the virtual hopping process. The pseudospin coupling along $z
$-direction includes both the exchange term $(\tau_i^+ \tau_j^- +h.c.)$ and
the pair flip term $(\tau_i^+ \tau_j^+ + h.c.)$. In spite of an isotropic
matrix in the $z$-direction, the orbital Hamiltonian is not $SU(2)$
symmetric because of the presence of Hund's coupling. In particular, the
pair flip term is related to superexchange processes involving those
intermediate states  listed in Table II that are split in energy due to
Hund's coupling. To illustrate this point further, we consider a pair of
V-ions along $z$-direction with antiparallel spins and pseudospins are $%
\tau_{iz}=\tau_{jz} =1/2$. The relevant intermediate states in the
superexchange are the states listed in the second and the fifth rows in
Table II. Because these states have different energies, there is non-zero
amplitude for the pseudospins to flip to $\tau_{iz}=\tau_{jz}=-1/2$. The
pseudospin pair flip process is actually quite common in orbital physics.
For example, there are pair flip terms in the effective Hamiltonian for
spin-1/2 systems with orbital degeneracy derived by Castellani et 
al.~\cite{castellani}.

In the limit $J/U=0$, $\eta _{\frac{3}{2}}=\eta _{\frac{1}{2}}=1$, 
and $\alpha =0$, and
so we have $K_{-}^{z}=Q^{2}$, and 
$K_{-}^{(x,y)}=-\frac{l_{x,y}}{2}(\tau _{iz}+\tau _{jz})+\frac{3}{2}$.
The orbital coupling between the two
V-ions of antiparallel spins vanishes, and the orbital coupling between the
two V-ions of parallel spins remains to be pseudospin $SU(2)$ symmetric
along $z$-direction and pseudospin Ising symmetric along $x$- or 
$y$-directions.
For $J/U=0$, the lack of the global pseudospin $SU(2)$ symmetry
is due to the anisotropic hopping integrals in the system.

Our effective Hamiltonian here is similar to  the Hamiltonian proposed
previously by Khaliullin et al.~\cite{giniyat} with the following
differences. These authors considered a model with 3-fold $t_{2g}$ orbital
degeneracy, while we consider a 2-fold orbital degeneracy with $d_{xy}$
orbital being always singly occupied. Below we shall compare the two
Hamiltonians by considering the the spin- orbital coupling between two
V-ions along $z$-direction, namely $H_{ij}$, with $j=i+z$. This may be
carried out by imposing the orbital $d_{xy}$ being always singly occupied in
the Hamiltonian of Khaliullin et al. We find that the $\tau _{iz}\tau _{jz}$
terms are the same in the two theories. However, the Hamiltonian of
Khaliullin et al. does not include the pseudospin flip term $(\tau
_{i}^{+}\tau _{j}^{+}+h.c.)$. As we illustrated above, the pseudospin flip
term is non-zero. As we shall see in the next section, this pair flip term
does not affect the mean field results, which depend only on the
z-components of pseudopsin in the present case. It will be an interesting
question to examine if the pseudospin flip term is important to the orbital
fluctuations.

\section{III. \, Mean field theory and the phase diagram}

We start with the classical solutions of $H_{eff}$. The Hamiltonian has a
global SU(2)symmetry in spin space, so that we can assume the spin ordering
along the z-direction. The Hamiltonian is invariant under the simultaneous
transformation of global $Z(2)$ (reversing orbitals at all sites) and a $90^0
$ rotation of the lattice about the $z$-axis. In general, we should consider
orbital ordering along an arbitrary orientation. However, for the present
problem, the orbital z-component terms are always larger or equal to the x-
or y-component terms in $K_{\pm }^{\nu }$ of Eq. (6). Therefore, we can
discuss the classical solutions by considering the z-component of the
orbital ordering only~\cite{delta}. In other words, the classical solutions
are the same as the Ising solutions in the present case. In Table I, we show
the energies per site in various classical states. Note that $%
\eta _{\frac{1}{2}}\leq 1$, and $\eta _{\frac{3}{2}}\geq 1$, where the
equality holds if and only if $J=0$. We consider below the case with
non-zero Hund's coupling $J>0$. In this case, the two states listed in Table
I with C-type antiferro-orbital (CO) configuration have higher energies.
Also, we can see that the G-type antiferromagnetic spin (GS) and G-type
antiferro-orbital (GO) phase has a higher energy than the CS-GO phase.
Therefore, the ground state is either ferromagnetic spin (FS) and GO, or
CS-GO. In both FS-GO and CS-GO phases, the orbital is antiparallel in all
three directions, favored by the combination of the symmetries in hopping
integrals (due to the cubic crystal symmetry) and the Hund's coupling. The
asymmetry between the spin configuration along the z-axis and in the x-y
plane is a result of the splitting of the $d_{xy}$ orbital level from the
other two $t_{2g}$ orbitals. As expected, the FS-GO phase is energetically
more favored at a larger J where the Hund's coupling dominates, and the
CS-GO phase is more favored at a smaller J. It may be helpful to understand
these two possible ground states by examining the following limiting cases
more explicitly. In the limit of large Hund's coupling, $\eta \rightarrow 1/3
$, the terms in the energy expression in Table I proportional to $\eta _{%
\frac{3}{2}}$ dominate. Hence the FS-GO phase has the lowest energy. In the
limit $J\rightarrow 0^{+}$, $\eta _{\frac{1}{2}}\rightarrow 1-0^{+}$, and $%
\eta _{\frac{3}{2}}\rightarrow 1+0^{+}$, so that the CS-GO phase has the
lowest energy.

It should be noted that although the $d_{xy}$ orbital is always singly occupied,
the virtual hopping of the $d_{xy}$ electron means that
our model is not identical to a model with one electron
occupying degenerate $d_{xz}$ and $d_{yz}$ orbitals. In that model,
for G-type orbital ordering, the ground state with non-zero Hund's coupling would 
always be ferromagnetic. This is
also true if we compare our model to a model with one electron occupying
triply degenerate $t_{2g}$ orbitals~\cite{kugel2,khaliullin}.

\begin{table*}[htb]
\caption{Mean field classical ground state energies for various phases. }
\begin{tabular}{|c|c|}
\hline
\textbf{Phase} & \textbf{\ Energy per site} \\ \hline
G-type spin and G-type orbital (GS-GO) & $-\frac{2}{3}\biggr( %
5+2Q^2+(1+Q^2)\eta_{\frac{3}{2}}+3\eta_{\frac{1}{2}}\biggr)$ \\ \hline
C-type spin and G-type orbital (CS-GO) & $-\frac{2}{3}\biggr( 5+(1+Q^2)\eta_{%
\frac{3}{2}}+3\eta_{\frac{1}{2}}\biggr)$ \\ \hline
G-type spin and C-type orbital (GS-CO) & $-\frac{1}{3}\biggr( 10+3Q^2+2\eta_{%
\frac{3}{2}}+3(2+Q^2)\eta_{\frac{1}{2}}\biggr)$ \\ \hline
C-type spin and C-type orbital (CS-CO) & $-\frac{2}{3}\biggr( 5+\eta_{\frac{3%
}{2}}+3\eta_{\frac{1}{2}}\biggr)$ \\ \hline
Ferro spin and G-type orbital (FS-GO) & $-2(1+Q^2)\eta_{\frac{3}{2}}$ \\
\hline
\end{tabular}
\end{table*}

We now discuss the finite temperature phases and their transitions. We
introduce three types of order parameters, namely the spin order parameter $%
m_{i}=\langle S_{iz}\rangle $, the orbital order parameter $r_{i}=\langle
\tau _{iz}\rangle $, and the spin-orbital order parameter $q_{i}=\langle
S_{iz}\tau _{iz}\rangle $. We shall consider the order parameters
corresponding to the FS-GO and CS-GO phases, since other ordered states are
not energetically favorable. In both the FS-GO and CS-GO phases, we divide
the lattice into two sublattices $A$ and $B$ accordingly. For the FS-GO
ordering, we consider $m_{i}=m$ for all the sites $i$, and $r_{i}=r$ and $%
q_{i}=q$ for $i$ at sublattice $A$ and $r_{i}=-r$ and $q_{i}=-q$ for $i$ at
sublattice $B$. For the CS-GO ordering, we consider $m_{i}=m$, $r_{i}=r$,
and $q_{i}=q$ for $i$ at sublattice $A$, and $m_{i}=-m$, $r_{i}=-r$, and $%
q_{i}=q$ for $i$ at sublattice $B$. We use a mean field theory to examine
the thermodynamically stable phases described by these order parameters, and
neglect both quantum and thermal fluctuations. The effective Hamiltonian $%
H_{eff}$ is then approximated by,

\begin{eqnarray}
H_{MF} = \sum_{i}(aS_{iz}+b\tau_{iz}+cS_{iz}\tau_{iz}+d)
\end{eqnarray}

\noindent In the above equations, coefficient $a,\,b,\,c,\,d$ are functions
of $\eta $ , $Q$, as well as of the mean fields $m$, $r$, and $q$. They are
given by, with the subscript upper (-) and lower (+) signs corresponding to
the CS-GO and FS-GO phases, respectively,
\begin{eqnarray}
a &=&A_{\mp}mr^2+ D_{\mp}m,  \nonumber \\
b &=& A_{\mp} m^{2}r- B_{\mp}r,  \nonumber \\
c &=& - A_{\mp}(2m^{2}r^{2}+q),  \nonumber \\
2d &=&-C_{\mp} m^{2}r^{2}+ E_{\mp} m^{2} + B r^2 - A_{\mp}q^{2},
\end{eqnarray}
where $B=B_{-}$, and
\begin{eqnarray}
A_{\pm }&=&4(\eta_{\frac{3}{2}}+ \alpha) (Q^2 \pm 1),  \nonumber \\
B_{\pm}& =&4(\eta_{\frac{3}{2}}-\alpha) (Q^2 \mp 1),  \nonumber \\
C_{\pm}&=&4(\eta_{\frac{3}{2}} +\alpha) (Q^2 \pm \frac{1}{2}), \\
D_{\pm} &=&(1+\eta_{\frac{1}{2}})(Q^2 \pm 3) - (\eta_{\frac{3}{2}} +
\alpha)(Q^2 \pm 1),  \nonumber \\
E_{\pm} &=& -(1+\eta_{\frac{1}{2}})(Q^2 \pm \frac{3}{2}) + (\eta_{\frac{3}{2}%
} + \alpha)(Q^2 \pm \frac{1}{2}).  \nonumber
\end{eqnarray}

The mean field Hamiltonian can be solved easily to obtain the thermal
averages of $S_{iz}$, $\tau_{iz}$, and $S_{iz}\tau_{iz}$, from which we
obtain the following self-consistent equations for the order parameters $m$,
$r$ and $q$, (with $\beta$ the inverse temperature)
\begin{eqnarray}
\langle S_{iz} \rangle &=& m = -\frac{2\sinh (\beta a)}{(1+2\cosh (\beta a))}
\nonumber \\
\langle \tau_{iz} \rangle &=& r = -\frac{1}{2}\tanh (\beta b/2)  \nonumber \\
\langle S_{iz}\tau_{iz} \rangle &=& q = - \frac{\sinh (\beta c/2)}{(1+2\cosh
(\beta c/2))}.
\end{eqnarray}
The free energy per site in the mean field theory is given by
\begin{eqnarray}
f &=& -\frac{1}{\beta}\ln \biggr(4\cosh(\beta b/2)[1+2\cosh(\beta c/2)]
\nonumber \\
&&\times [1+2\cosh(\beta a)]\biggr) + d.
\end{eqnarray}

We solve the self-consistent equations for different ordered states at
various temperatures. The phases studied are (i). paramagnetic spin (PS) and
para-orbital (PO) state (PS-PO) with $m=r=q=0$; (ii). CS-PO state with
C-type spin ordering $m\neq 0$ and $r=q=0$; (iii). PS-GO state with $r\neq 0$
and $m=q=0$; (iv). CS-GO state; and (v). FS-GO state. In the states (iv) and
(v), $m,r,q\neq 0$. When more than one set of MF solutions exist at a given
temperature, we compare their free energies to determine the
thermodynamically stable phase diagram.

\begin{figure}[htb]
\epsfxsize=5.1cm \centerline{\epsffile{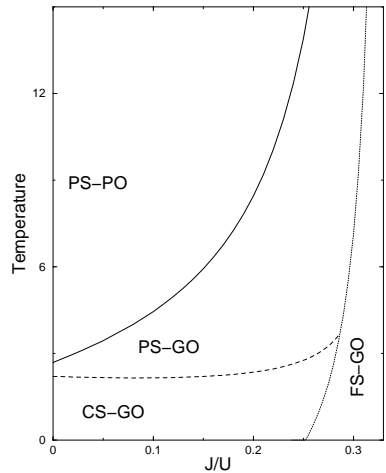}}
\caption{Phase diagram for the anisotropic hopping parameter $%
Q=t_{\perp}/t_{\parallel} =1.3$ in the parameter space of temperature and
relative Hund's coupling $J/U$. Temperature is shown in units of $%
t_{\parallel}^{2}/U$. Solid line is the phase boundary between para-orbital
(PO) and G-type antiferro-orbital (GO) states, the dashed line is the
boundary between paramagnetic (PS) and C-type antiferromagnetic (CS) states,
the dotted line is the phase boundary between PS and ferromagnetic (FS)
states.}
\label{}
\end{figure}

\begin{figure}[htb]
\epsfxsize=5.5cm \centerline{\epsffile{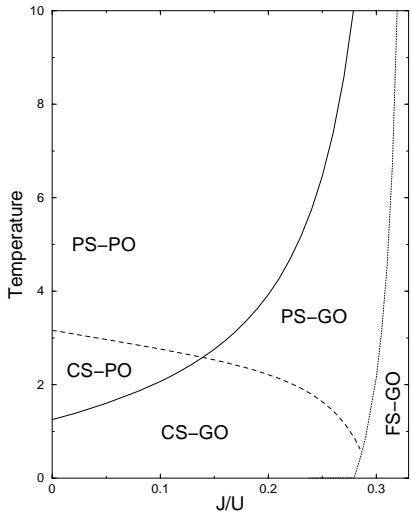}}
\caption{Phase diagram for $Q =t_{\perp}/t_{\parallel} = 0.5$. For details,
see Fig.2 caption. }
\label{}
\end{figure}

In Fig. 2 and Fig. 3, we plot the phase diagrams obtained from the mean
field theory for $Q=1.3$ and $Q=0.5$, respectively. The phase diagram for $%
Q=1$ is qualitatively the same as that for $Q=1.3$. The ground state is
found to be CS-GO for smaller J/U, and FS-GO for larger J/U, consistent with
our previous discussions. In general, the spin and orbital ordering occur at
different temperatures. This feature is in distinction from the model for $%
V_{2}O_{3}$ for which spin and orbital order at the same temperature~\cite
{mila,joshi}. Within the mean field theory, the phase transition between
CS-GO and FS-GO is first order, and all other transitions between the
different phases in Figs. 2 and 3 are second order. (The lattice distortion
associated with the orbital ordering, which is not included in our model,
may change the nature of the phase transition.)

In passing, we note that the spin-orbital ordering described by the order
parameter $q$ is always zero unless both the spins and the orbitals are
ordered in the present system. This result indicates that the spin-orbital
ordering parameter $q$ introduced in our mean field theory in addition to
the spin order parameter $m$ and the orbital order parameter $r$ may not be
as significant in the present problem in altering the qualitative physics as
in the $SU(4)$ model~\cite{li}.

In what follows, we shall focus on the phases relevant to $YVO_{3}$, and
discuss the sequential phase transitions from PS-PO to CS-GO. As we can see
from Fig. 2, as the temperature decreases from the disordered state PO-PS,
the system first undergoes a transition to the G-type antiferro-orbital
ordered phase. Only at lower temperatures does the spin become C-type
antiferromagnetic ordered. For smaller $Q$, the phase transitions depend on
the Hund's coupling as we can see from Fig. 3. At intermediate $J/U$, the
orbital transition temperature is higher than the spin's, while at smaller $%
J/U$, the spin transition temperature is higher than the orbital's. Since
for $YVO_{3}$, $Q\geq 1$, and the estimated values for $U$ and $J$ are $%
U\sim 4.5eV$ and $J\sim 0.68eV$~\cite{mizokawa2,sawada}, our theory suggests
orbital ordering first at a higher temperature followed by a subsequent spin
ordering at a lower temperature in $YVO_{3}$. This is qualitatively
consistent with the experimental findings for $YVO_{3}$ above 77K.

As recently reported by Blake $\mathit{et.al}$ ~\cite{blake}, the neutron
diffraction experiment shows that the orbital ordering in $YVO_3$ takes
place at $T_{GO}=200$K, which is far above the antiferromagnetic ordering
temperature $T_{CS}=116$K. The orbital ordering is evidenced by the changes
of the V-O bond lengths in the xy-plane. Our theory is consistent with these
observations. Orbital ordering and lattice distortion are often observed
simultaneously in experiments. It is usually difficult to distinguish if the
lattice distortion is due to the orbital ordering or vice versa. Our theory
suggests a scenario, in which the orbital ordering is of electronic origin,
and the lattice distortion observed above 77K is a consequence of the
orbital ordering and the electron-phonon coupling.

In $YVO_{3}$, as temperature decreases further, there is another phase
transition at a lower temperature $T_{GS}=77$K, below which the system is in
the G-type antiferromagnetic and C-type antiferro-orbital state. There have
been proposals~\cite{mizokawa2,giniyat} to attribute this lower temperature
phase transition to the Jahn-Teller energy which favors C-type orbital
ordering. It is an interesting issue to further understand the nature of the
low temperature phase transition.

\section{IV. \, Summary}

In summary, we have studied the electronic structure of the insulating $%
YVO_{3}$, and derived an effective Hamiltonian based on the superexchange
interaction. We started with the atomic limit where each $V-3d^{2}$ has a
spin-1 and two-fold degenerate orbital configurations $(d_{xy}d_{xz})$ and $%
(d_{xy}d_{yz})$. This consideration is consistent with the recent neutron
diffraction experiment at $T>77$K. We studied the classical solutions of the
model within mean field theory, and found G-type antiferro-orbital ordering
at a higher temperature followed by a second phase transition where the
spins become C-type ordered. Our theory explains the orbital and spin
ordering of $YVO_{3}$ at temperatures $T>77$K. While our model does not
explain the lower temperature phase of G-type spin ordering, which may
require considerations in addition to the superexchange interaction, our
theory provides a starting point for understanding the unusual magnetic
properties of $YVO_{3}$.

After we completed the present calculations, we learned of a very recent
inelastic neutron scattering experiment of Ulrich et al.~\cite{keimer}. They
reported an energy gap in the spin wave spectrum of $YVO_{3}$ in the C-type
spin ordering phase, and they interpreted it as an evidence for the orbital
Peierls state along the z-direction. The classical solutions we study here
do not predict any spin or orbital Peierls transition. Quantum fluctuations
or electron-lattice interactions may be responsible for this unusual state~%
\cite{keimer,shen}.

This work was in part supported by NSF Grant \#0113574, the URC Summer
Student Fellowship at University of Cincinnati, and by the Chinese Academy of
sciences. MM acknowledges the hospitality of the Hong Kong University of
Science and Technology.

\appendix

\section{V. \, Appendix}

In this Appendix, we present all the intermediate eigenstates $| I \rangle$
and their corresponding energy differences with the ground state, $E_I - E_0$%
, of the the four $t_{2g}$ electrons in two V-ions. These states are used in
the second order perturbation theory to derive the effective Hamiltonian $%
H_{eff}$ in the text. The atomic ground states of the system are $6 \times 6$
-fold degenerate, with each ion occupied by two electrons of parallel spins
in the orbital configurations $(d_{xy}d_{xz})$ or $(d_{xy}d_{yz})$. The
atomic ground state energy is $E_0 = 2 (U-3J)$. The excited states $| I
\rangle$ are $(V-3d^1)-(V-3d^3)$ with 6-fold degeneracy in $V-3d^1$ ion.
Here we consider the limiting case $U, \, J >> \Delta$, and neglect the
effect of $\Delta$ on the $V-3d^3$ states. Within this approximation, these
excited states are split into three multiples due to the Hund's coupling: $%
4\times 6$ states with energy $(U-3J+E_0)$, $10 \times 6$ states with energy
$(U+E_0)$, and $6 \times 6$ states with energy $(U+2J+E_0)$. The spin and
orbital configurations of $V-3d^3$ are listed in Table II.

\begin{table*}[htb]
\caption{Intermediate eigenstates $| I \rangle$ and corresponding energy
differences $E_I - E_0$. The degeneracy shown is respected with the orbital
configurations.}
\begin{tabular}{|l|c|r|}
\hline
\textbf{Eigenstate} & \textbf{Spin and orbital configuration} & \textbf{$E_I
- E_0$} \\ \hline
$(c^{\dagger}_{1\uparrow}c^{\dagger}_ {2\uparrow}c^{\dagger}_
{3\uparrow})|0\rangle$ & \epsfysize=0.8cm \epsffile{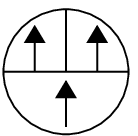} &  \\
$\frac{1}{\sqrt{3}}(c^{\dagger}_{1\uparrow}c^{\dagger}_{2\downarrow}c^{%
\dagger}_ {3\uparrow}+c^{\dagger}_{1\downarrow}c^{\dagger}_
{2\uparrow}c^{\dagger}_ {3\uparrow}+c^{\dagger}_{1\uparrow}c^{\dagger}_
{2\uparrow}c^{\dagger}_ {3\downarrow})|0\rangle$ & \epsfysize=0.8cm %
\epsffile{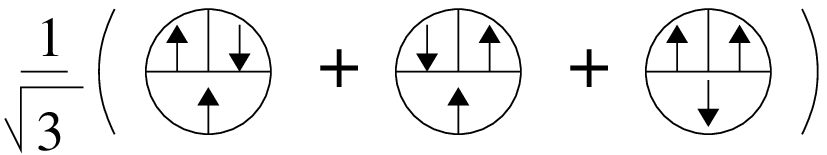} & $U-3J$ \\
$\frac{1}{\sqrt{3}}(c^{\dagger}_{1\downarrow}c^{\dagger}_{2\uparrow}c^{%
\dagger}_{3\downarrow}+c^{\dagger}_{1\uparrow}c^{\dagger}_{2\downarrow}c^{%
\dagger}_{3\downarrow}+c^{\dagger}_{1\downarrow}c^{\dagger}_{2\downarrow}c^{%
\dagger}_{3\uparrow})|0\rangle$ & \epsfysize=0.8cm \epsffile{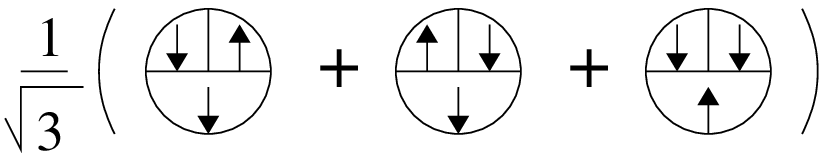} &  \\
$(c^{\dagger}_{1\downarrow}c^{\dagger}_ {2\downarrow}c^{\dagger}_
{3\downarrow})|0\rangle$ & \epsfysize=0.8cm \epsffile{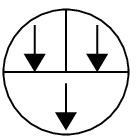} &  \\ \hline
$\frac{1}{\sqrt{2}}(c^{\dagger}_{1\uparrow}c^{\dagger}_
{1\downarrow}c^{\dagger}_ {3\uparrow}-c^{\dagger}_{2\uparrow}c^{\dagger}_
{2\downarrow}c^{\dagger}_ {3\uparrow})|0\rangle$ & \epsfysize=0.8cm %
\epsffile{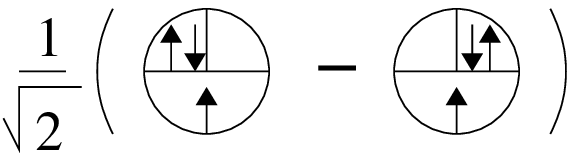} & $U$ \\
$\frac{1}{\sqrt{2}}(c^{\dagger}_{1\uparrow}c^{\dagger}_{1\downarrow}c^{%
\dagger}_ {3\downarrow}-c^{\dagger}_{2\uparrow}c^{\dagger}_
{2\downarrow}c^{\dagger}_ {3\downarrow})|0\rangle$ & \epsfysize=0.8cm
\epsffile{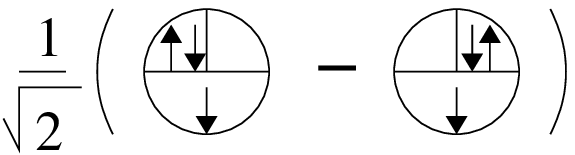} & (3-fold) \\ \hline
$\frac{1}{\sqrt{6}}(2c^{\dagger}_{1\uparrow}c^{\dagger}_
{2\uparrow}c^{\dagger}_ {3\downarrow}-c^{\dagger}_{1\uparrow}c^{\dagger}_
{2\downarrow}c^{\dagger}_ {3\uparrow}-c^{\dagger}_{1\downarrow}c^{\dagger}_
{2\uparrow}c^{\dagger}_ {3\uparrow})|0\rangle$ & \epsfysize=0.8cm %
\epsffile{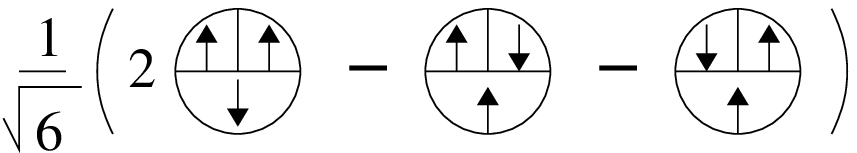} & $U$ \\
$\frac{1}{\sqrt{6}}(2c^{\dagger}_{1\downarrow}c^{\dagger}_{2\downarrow}c^{%
\dagger}_ {3\uparrow}-c^{\dagger}_{1\downarrow}c^{\dagger}_
{2\uparrow}c^{\dagger}_ {3\downarrow}-c^{\dagger}_{1\uparrow}c^{\dagger}_
{2\downarrow}c^{\dagger}_ {3\downarrow})|0\rangle$ & \epsfysize=0.8cm
\epsffile{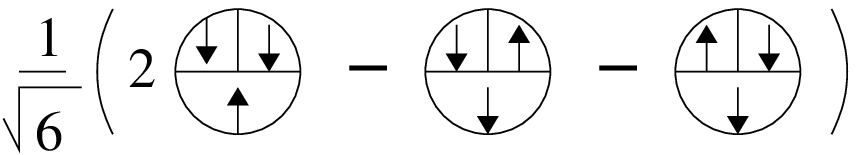} &  \\ \hline
$\frac{1}{\sqrt{2}}(c^{\dagger}_{1\uparrow}c^{\dagger}_
{2\downarrow}c^{\dagger}_ {3\uparrow}-c^{\dagger}_{1\downarrow}c^{\dagger}_
{2\uparrow}c^{\dagger}_ {3\uparrow})|0\rangle$ & \epsfysize=0.8cm
\epsffile{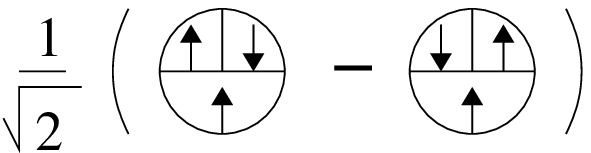} & $U$ \\
$\frac{1}{\sqrt{2}}(c^{\dagger}_{1\uparrow}c^{\dagger}_{2\downarrow}c^{%
\dagger}_ {3\downarrow}-c^{\dagger}_{1\downarrow}c^{\dagger}_
{2\uparrow}c^{\dagger}_ {3\downarrow})|0\rangle$ & \epsfysize=0.8cm %
\epsffile{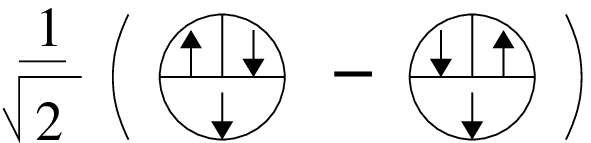} &  \\ \hline
$\frac{1}{\sqrt{2}}(c^{\dagger}_{1\uparrow}c^{\dagger}_
{1\downarrow}c^{\dagger}_ {3\uparrow}+c^{\dagger}_{2\uparrow}c^{\dagger}_
{2\downarrow}c^{\dagger}_ {3\uparrow})|0\rangle$ & \epsfysize=0.8cm %
\epsffile{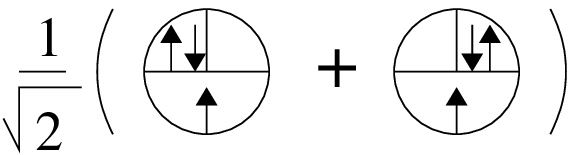} & $U+2J$ \\
$\frac{1}{\sqrt{2}}(c^{\dagger}_{1\uparrow}c^{\dagger}_{1\downarrow}c^{%
\dagger}_ {3\downarrow}+c^{\dagger}_{2\downarrow}c^{\dagger}_
{2\uparrow}c^{\dagger}_ {3\downarrow})|0\rangle$ & \epsfysize=0.8cm
\epsffile{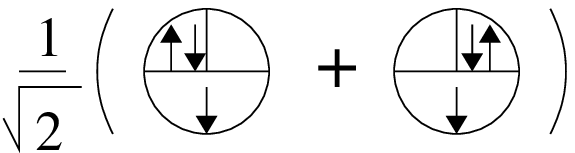} & (3-fold) \\ \hline
\end{tabular}
\par
\begin{center}
\end{center}
\end{table*}


\begin{references}
\bibitem{delta} We have examined the linear orbital operator term in
Eq. (6).  For the spin configurations considered in this paper, namely
the C- or G-type antiferromagnets or ferromagnet,
the contribution of this linear term to the energy vanishes.
\end{references}
\end{document}